\documentstyle[preprint,aps,floats,psfig]{revtex}
\begin{document}

\draft

\title{Anharmonic decay of vibrational states in amorphous silicon} 
\author{Jaroslav Fabian and Philip B. Allen}
\address{Department of Physics, State University of New York at Stony Brook, 
Stony Brook, New York 11794-3800}
\maketitle

\vspace{1cm}

\begin{abstract}

Anharmonic decay rates are calculated for a realistic atomic model
of amorphous silicon. The results show that the vibrational 
states decay on picosecond timescales and follow the two-mode
density of states, similar to crystalline silicon, but somewhat faster.
Surprisingly little change occurs for localized states. These results
disagree with a recent experiment.

\end{abstract}
\vspace{2em}
\pacs{PACS numbers: 63.20.-e, 63.20.Kr, 63.20.Pw, 63.50.+x}

\newpage

Scholten et al.\cite{scholten1,scholten2} have  found long
relaxation times
of vibrational states (VS) in amorphous silicon (a-Si) 
and hydrogenated amorphous silicon (a-H:Si), at liquid He
temperatures. Hot charge carriers produced by laser pulses
excite VS during their thermal relaxation towards 
the conduction band edges and the time dependence of the VS populations
is measured by anti-Stokes Raman spectroscopy.
Their main observations are (a) the populations of high frequency VS 
live about 10 ns, and (b) the relaxation times increase with frequency.
Similar results have been reported for Cr$^{3+}$ doped borate
glass using far infra-red radiation for the VS excitation \cite{exp}.
The situation in crystals is very different: optical phonons decay 
on picosecond timescales and the lifetimes of phonons typically 
decrease with frequency, as new decay channels become available.

The model of Orbach et al.\cite{orbach1,orbach2} offers an
explanation of this controversy. 
It assumes that most of the VS are localized (fractons,  since
the localization is implemented for a glass with fractal structure). 
The mobility edge occurs at low frequencies ($\approx$ 5 meV) and 
is equated with the Ioffe-Regel limit (IRL), where the wavevector 
is no longer a quantum number.  
The only relevant decay channel for the high frequency fractons
is  decay to another fracton and a low frequency phonon
(hopping). The decay to two phonons is kinematically
forbidden, and the probability of a three-fracton interaction is
considered to be small. The above two experimental observations
can then be rationalized \cite{scholten1,orbach2}. However,
experiment provides no indication of fractal structure
in homogeneous glasses like a-Si.

As an alternative model, Allen and Feldman used the
concept of extended and non-propagating modes \cite {allen1,allen2}
to explain the thermal conductivity of a-Si
at high temperatures. Numerical simulations \cite{allen2,local} show an
onset of localization far beyond the IRL.
The majority of modes have frequencies between the IRL and the
mobility edge, and are extended but not propagating.

So far only 
harmonic vibrational properties of disordered structures have
been systematically studied, but it is
natural, in view of the above experiments, to include
anharmonic couplings and investigate perturbatively 
the decay properties of different VS. 
The situation in 
crystalline silicon (c-Si) is well understood both 
experimentally \cite{temple,cardona} and theoretically 
\cite{vanderbilt,baroni}, so the comparison between the 
crystalline and amorphous cases provides
another motivation for the present calculations.

In this Letter we show that contrary to the experimental results 
the VS in a realistic model of a-Si decay on picosecond scales 
and their lifetimes decrease with frequency, as in c-Si.  
First we calculate the linewidths of the zone-center 
optical phonons in c-Si, using the Stillinger-Weber (SW) 
interatomic potential 
\cite{stillinger}. Next we use the same potential to study the decay 
rates of VS in a model amorphous system of 216 Si atoms. 
The decay rates of VS in a hypothetical structure
of amorphous Si/Ge with the germanium mass increased by the factor
of two are then computed and the results are discussed.

It has been known that the SW  potential, which consists of the two- 
and three-body terms, 
reproduces reasonably well elastic properties of different phases
of silicon \cite{li2}. Vibrational spectra have been calculated for both
crystalline \cite{li2,li1} and amorphous silicon \cite{allen2,li2}, and give 
reasonable agreement with 
experiment, overestimating by 15\% the high frequency region.  
The high temperature value of the thermal
expansivity \cite{li2} and the values of pressure derivative of elastic 
constants \cite{feldman} are underestimated by the SW potential, indicating 
its relatively small anharmonicity.

The decay rate (inverse lifetime) $2\Gamma(j)$ 
of the mode j with the frequency $\omega(j)$, at a temperature T, 
is given by \cite{maradudin}
\begin{eqnarray} \nonumber
2\Gamma(j)=\frac{\hbar^2\pi}{4\omega(j)}\sum_{k,l}
\frac{\left| V(j,k,l) \right| ^2} 
{\omega(k)\omega(l)}(\frac{1}{2}[1+n(k)+n(l)]  
\delta[\omega(j)-\omega(k)-\omega(l)] \\ \label{eqn:decay}
+ [n(k)-n(l)]\delta[\omega(j)+\omega(k)-\omega(l)]) 
\end{eqnarray}
Here V(j,k,l) are the coefficients of the cubic terms in the expansion
of the interatomic potential in normal mode coordinates, and n(k) are
the Bose-Einstein occupation factors $[\rm exp(\hbar\omega(k)/kT) - 1]^{-1}$.
Due to lack of translational invariance, the VS in amorphous 
systems can be characterized according to their frequency only. 
For phonons in crystals, j denotes both the band index $\rm \lambda$ and 
the wavevector {\bf q}, and the wavevector conservation up to a reciprocal 
lattice vector is a selection rule in V(j,k,l).
Two different physical processes lead  to the above equation:
(a) decay of the mode j to all energy conserving
combinations of k and l with the temperature factor 1+n(l)+n(k),
and (b) decay when the mode j is destroyed together 
with the mode k (and the mode l is created) has the temperature
factor n(k)-n(l). 
Delta functions ensure energy conservation. 
Note that at zero temperature only the (a) decay exists and at
high temperatures the decay grows proportionally to the temperature.

As a test for the SW potential we have carried out the calculation of 
the decay rates
of the zone-center optical phonons in c-Si. In this case only processes (a)
are contributing (frequencies of the {\bf q}={\bf 0} phonons
are the highest in the vibrational spectrum). By identifying the
decay rates with the full widths at half maxima of Raman absorption
lines, one can also make comparison with experiment. The sum over the
{\bf q} points in Eq. (1) is performed by the tetrahedron method,
using 1772 tetrahedra in the irreducible wedge of the Brillouin
zone. Figure \ref{fig:1} shows the temperature dependence of the
decay rate along with the experimental values from Ref. \cite{cardona}.
At zero temperature our calculated decay rate is 0.064 meV,
as compared to 0.153 meV obtained experimentally \cite{cardona},
and 0.059 meV \cite{vanderbilt} and 0.183 meV \cite{baroni} obtained
using the local density approximation.
In accord with the results
of the thermal expansion \cite{li2} and pressure dependence of elastic 
constants \cite{feldman},
we find the SW potential underestimating anharmonic
effects. The dominant decay mechanism involves one LA and one TA mode
($\approx$ 93\%), while the diagonal LA + LA Klemens channel \cite{klemens}
gives a relatively small contribution ($\approx$ 7\%). This follows from
the trends in the vibrational density of states, as observed earlier
\cite{vanderbilt,baroni}.

To study the decay rates in a-Si we represent its structure
by a cubic unit cell
of side 16.35 $\AA$, containing 216 atoms (27 fcc unit cells),
and periodic boundary conditions. The random network model atomic coordinates
by Wooten, Winer, and Weaire \cite{wooten}, are further relaxed to a local
minimum of the SW potential \cite{li2}.
We use the same potential for the interatomic interactions. 
Harmonic vibrational properties of the model have been reported
elsewhere \cite{allen2}. Only about 3\% of VS above the mobility edge
$\omega_c\approx$ 71 meV are localized and most of the modes are extended.
Only modes with $\omega \alt$ 15 meV are marginally propagating
(partial memory of a wavelength remains).
The lowest frequency modes ($\alt$ 10 meV) are affected by finite size 
of our sample and are omitted from our calculations.

Due to the discreteness of the vibrational energy levels in our finite system, 
we represent the delta functions in the Eq. \ref{eqn:decay} by normalized
rectangles of the widths 0.4 meV, centered at the VS frequencies $\omega(j)$. 
On average there are 3.6 modes in a rectangle. Increase of the rectangle 
widths increases the computation time without a notable change in 
results. Smaller values enhance the noise. 
In Fig.\ref{fig:2} we show the calculated decay 
rates as frequency functions at 10K and 300K. The curves were convolved
with rectangles to suppress the remaining noise. 
The decay rates 2$\Gamma(j)$ are of order 0.01 $\omega (j)$, which
confirms the applicability of lowest order perturbation theory.
For lifetimes
one thus obtains picoseconds in contrast to the experimentally claimed
nanoseconds ($\hbar$/1 meV $\approx$ 0.7 ps). The decay rate follows the 
joint two-mode density
of states ($\sum_{j,k} \delta[\omega-\omega(j)-\omega(k)]$) curve, even 
beyond the mobility edge (indicated by the
vertical line). Since the localized states in our system 
kinematically decay mostly to two extended states, the decay properties 
of the
localized states resemble those of the extended states.
At higher temperatures the mid-frequency region becomes more
pronounced due to the appearence of processes (b).
The high temperature shape of the decay rate frequency dependence 
can be shown to follow a combination of the joint and difference 
two-mode density of 
states. Figure \ref{fig:3} provides information about the distribution of
the anharmonic matrix elements V(j,k,l), corresponding to the 
decay of the extended mode j to modes k and l. To improve statistics we
combined the information from three (randomly chosen) j modes with
frequencies between 65 and 68 meV. Since there are no selection rules
present for the decay of the extended modes (other than the energy
conservation), all kinematically allowed final states contribute on
average equally to the decay. Individual matrix elements occur with a
Gaussian distribution reflecting the disorder present in the amorphous 
structure, and the complete destruction of wavevector selection rules.

We now ask what changes would occur in decay rates if a greater portion
of the states were localized. To accomplish this we model a
hypothetical alloy of silicon and heavy germanium (a-Si/Ge) by changing 
the masses
of 50\% of the atoms chosen at random, to twice the germanium mass. 
The interatomic potential is unchanged. Harmonic vibrational properties of
this systems were studied \cite{allen2}, and it was shown that the mobility 
edge
shifts down to 31 meV. The extrinsic mass disorder increases
the portion of localized states. Figure \ref{fig:4} shows the decay rate
frequency functions at two temperatures, 10 K and 300 K. The mobility
edge is indicated by the vertical line. Again, there are no dramatic
changes beyond the mobility edge, the curves basically follow
the trends in the joint, and  the joint plus the difference 
two-mode density of states.
Modes with frequencies above 62 meV are like the modes of
the fracton model, as they
decay kinematically only via hopping; allowed decays
through three-localized mode interaction are of neglibible probability.
In order for a localized mode to decay to another localized mode and
an extended one, the two localized modes must overlap. As the localization
length of the localized modes decreases with frequency, the overlap probability
decreases. The deficiency in the number of hopping decay channels
is in our model compensated by the increase in magnitude of the hopping matrix
elements (squares of the vibrational amplitudes scale inversely with the 
localization length). We therefore find similar decay rates for the 
localized and
extended states, although their decay mechanisms are very different.

Compared with the decay of the optical phonons in c-Si,
the decay rates of the a-Si VS just below the mobility edge
are larger by a factor of three. Considering the bond-angle and 
other structural disorder in a-Si, this seems reasonable. However, 
the spectral characteristics of the final decay products are very 
different for the two cases.
All kinematically allowed decay channels contribute random
terms in the a-Si case (Fig.\ref{fig:3}), while due to translational 
invariance there is only a small spectral region of allowed decay
products in the c-Si case \cite{vanderbilt,baroni}. As in the case
of hopping, the deficiency is compensated by the larger magnitudes
of the decay matrix elements in c-Si.

In conclusion, we have found that the high-frequency VS in a realistic model
of a-Si decay on picosecond time scales, and at low temperatures their 
lifetimes decrease as frequency increases. This is in contrast to recent
experimental claims that the modes decay on nanosecond scales and
their lifetimes increase as frequency increases. 
Two remarks about the experiment are
appropriate. First, the average (non-equilibrium) population of the 
excited VS is $\rm \overline{n}(j) \approx$ 0.2 and the VS are excited 
over the whole spectrum \cite{scholten2}. However, decay rates 
$2\Gamma(j)$, Eq. \ref{eqn:decay}, are defined 
for {\it small}
deviations of {\it single-frequency} VS populations from equilibrium.
It is not clear that the experiment measures the same rates.
Second, the VS population is excited through the relaxation of hot
carriers. In amorphous systems carriers trapped in  
localized states in band tails can decay on timescales 
much larger than picoseconds and replenish the VS population.
It is possible that the experiment probes the slow relaxation of 
the coupled system of trapped carriers and VS, and not the intrinsic 
VS decay.

We thank J. L. Feldman and S. Bickham for useful discussions. 
J. F. thanks J. L. 
Feldman for hospitality at Naval Research Laboratory. This work
was supported by NSF Grant No. DMR 9417755.

\newpage

\begin{figure}
\caption{Decay rate of the zone-center optical phonons in c-Si from the
SW potential as a
function of absolute temperature. The filled circles are the experimental
points from Ref. 10.
}
\label{fig:1}
\end{figure}

\begin{figure}
\caption{Anharmonic decay rate of VS in a-Si as a frequency function
for temperatures 10 K and 300 K. The dashed line in the low temperature
graph is the joint two-mode density of states in arbitrary units
and the vertical line indicates the mobility edge $\omega_c\approx$ 71 meV.
}
\label{fig:2}
\end{figure}

\begin{figure}
\caption{(a) Distribution of the anharmonic matrix elements V(j,k,l) for
the decay of a sample of three extended normal modes j of
$\omega(j)$ between 65 and 68 meV.
The horizontal frequency axis
is the lower frequency of the two final states k and l; the other frequency
is given by the energy conservation $\omega(j)=\omega(k)+\omega(l)$.
(b) The number of occurences of the anharmonic matrix elements from (a) shows
their Gaussian random character due to lack of selection rules
other than the energy conservation.
}
\label{fig:3}
\end{figure}

\begin{figure}
\caption{Anharmonic decay rate of VS in hypothetical (see text)
a-Si/Ge as a frequency function
for temperatures 10 K and 300 K. The vertical line
indicates the mobility edge $\omega_c\approx$ 31 meV.
}
\label{fig:4}
\end{figure}

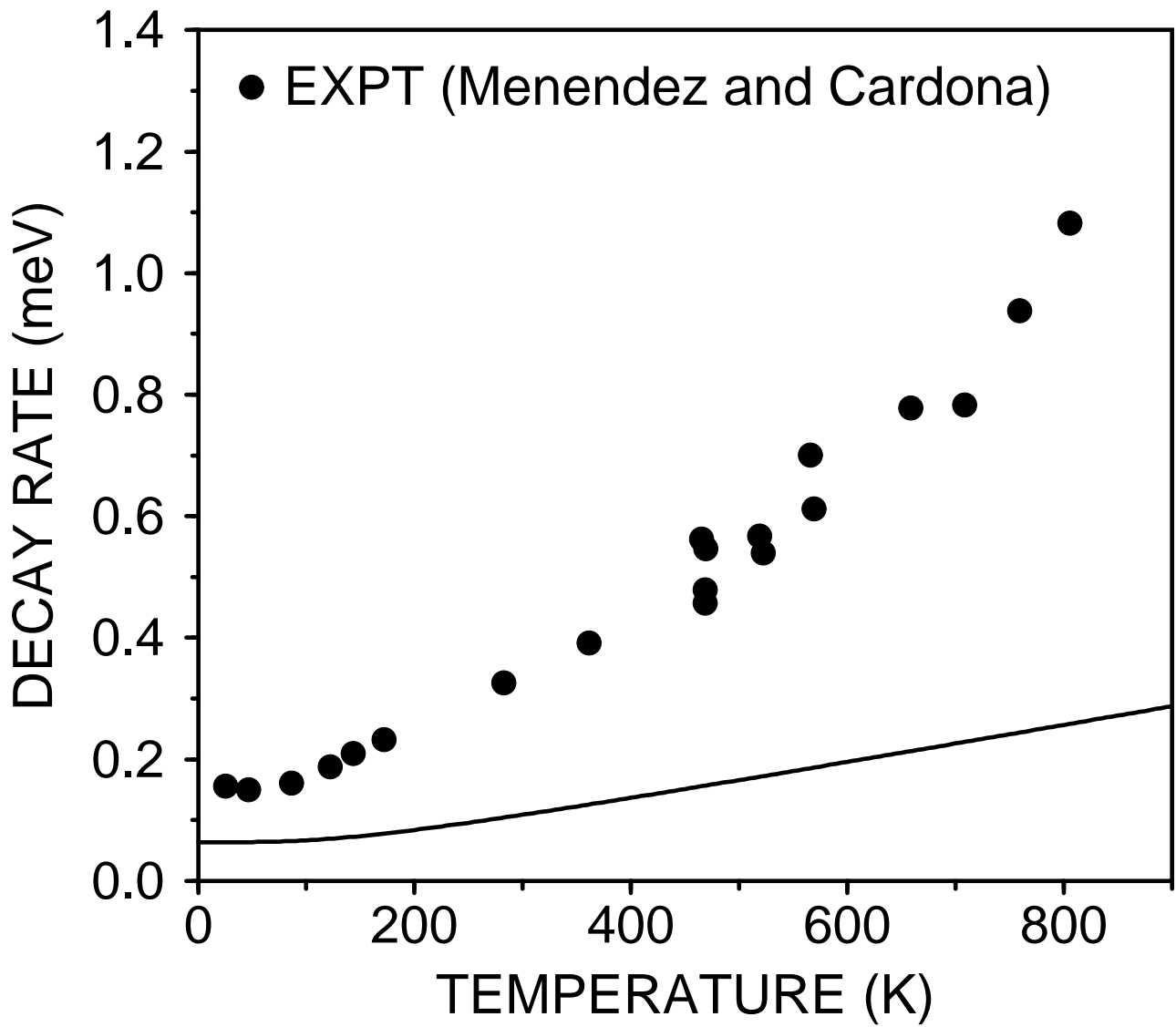
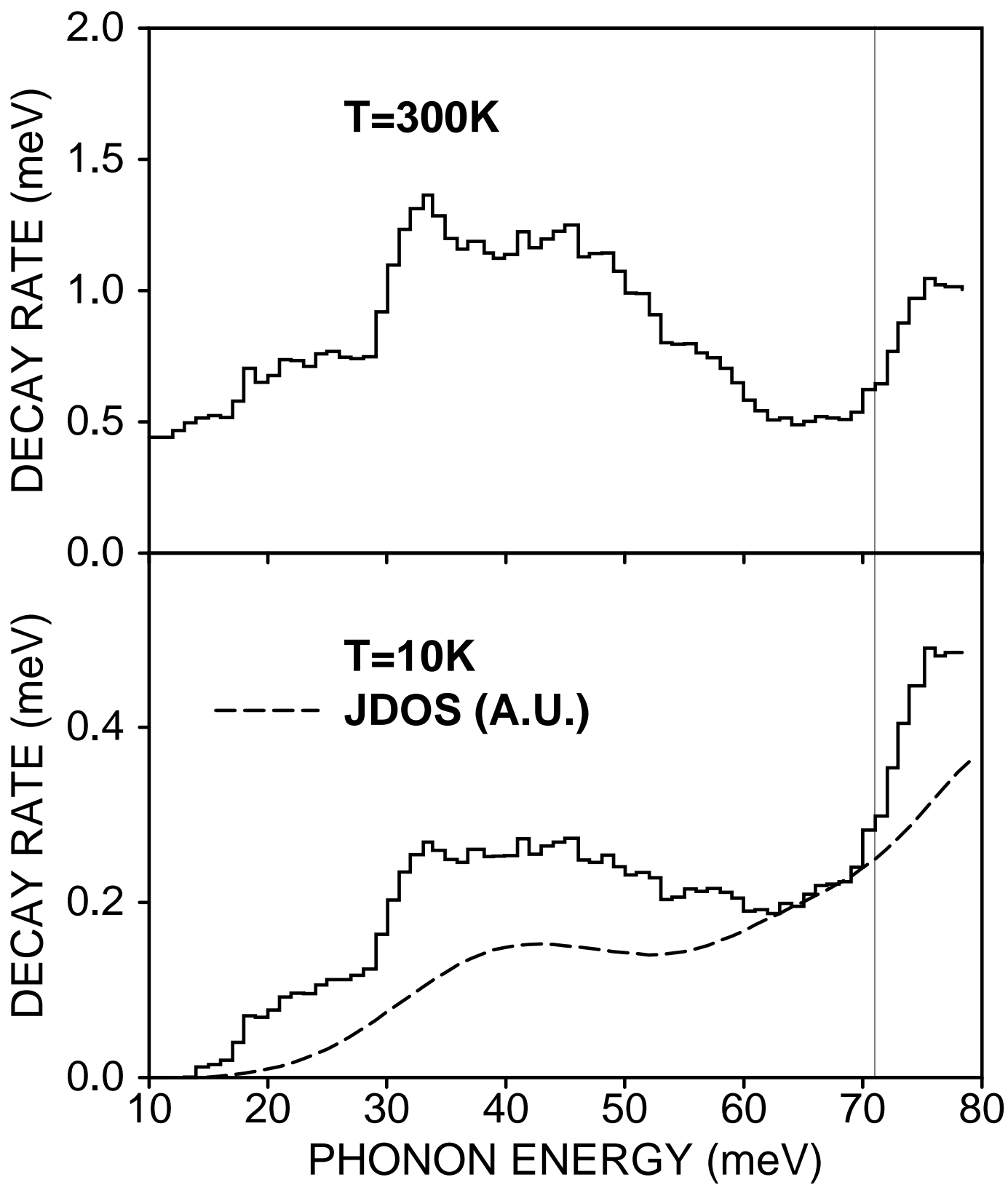
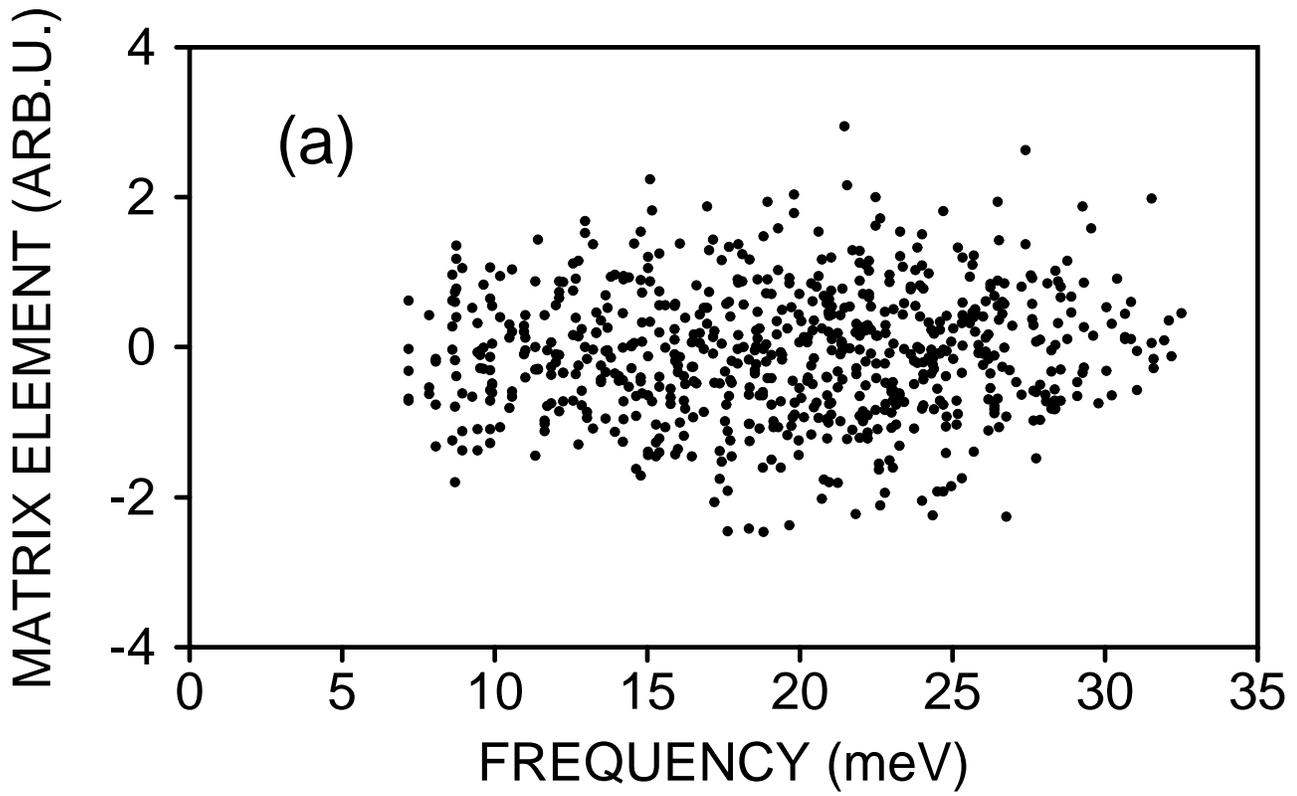
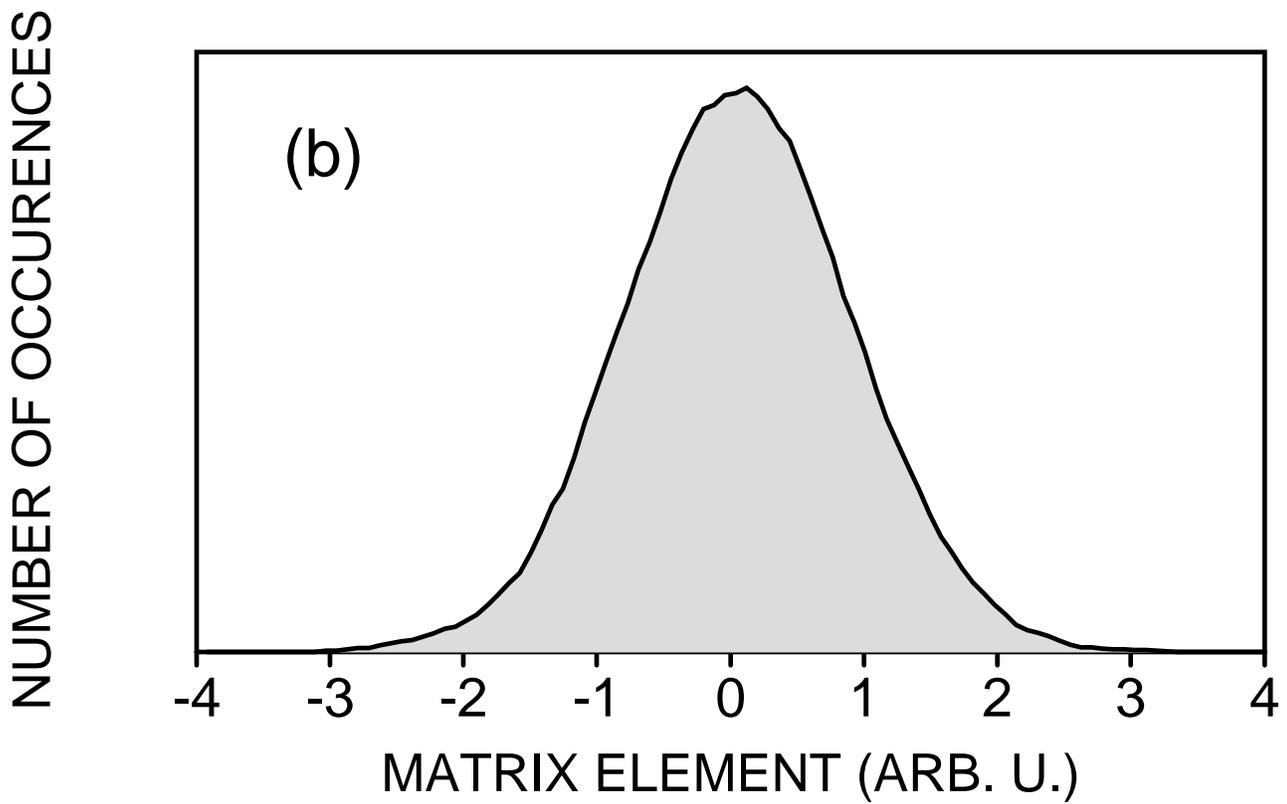
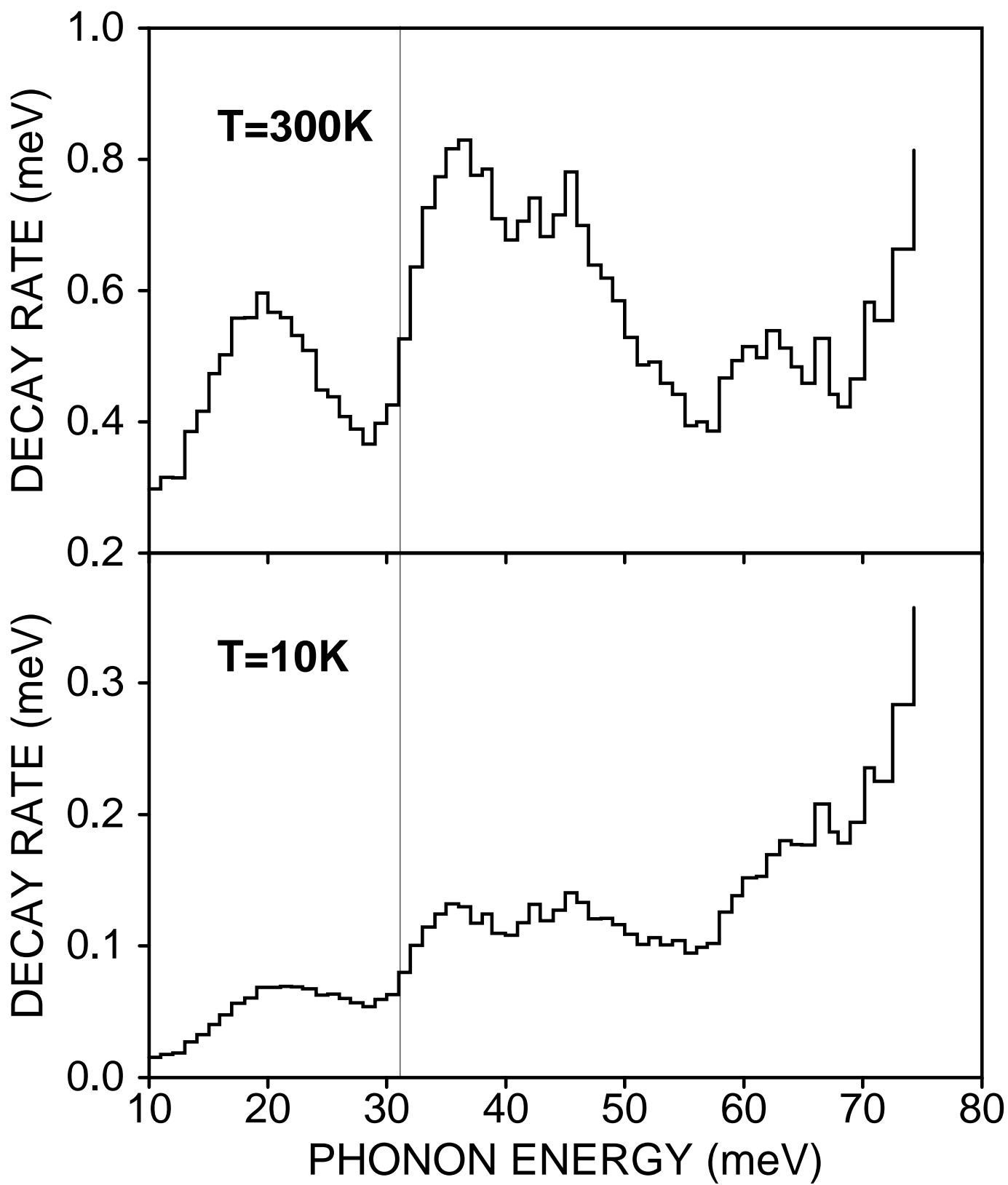
\end{document}